\documentclass[12pt]{article}
\usepackage[english,german,french,polish]{babel}
\usepackage[T1]{fontenc}
\usepackage{amsfonts}

\begin{document}

\selectlanguage{english}

\baselineskip 0.76cm
\topmargin -0.6in
\oddsidemargin -0.1in

\let\ni=\noindent

\renewcommand{\thefootnote}{\fnsymbol{footnote}}

\newcommand{\SM}{Standard Model }

\pagestyle {plain}

\setcounter{page}{1}

\pagestyle{empty}

~~~

\begin{flushright}
IFT-- 06/2
\end{flushright}

\vspace{0.4cm}

{\large\centerline{\bf A universal shape of empirical mass formula}}

{\large\centerline{\bf for all leptons and quarks{\footnote{Work supported in part by the Polish Ministry of Education and Science, grant 1 PO3B 099 29 (2005~-~2007). }}}} 

\vspace{0.4cm}

{\centerline {\sc Wojciech Kr\'{o}likowski}}

\vspace{0.3cm}

{\centerline {\it Institute of Theoretical Physics, Warsaw University}}

{\centerline {\it Ho\.{z}a 69,~~PL--00--681 Warszawa, ~Poland}}

\vspace{0.5cm}

{\centerline{\bf Abstract}}

\vspace{0.2cm}

A specific {\it universal} shape of empirical mass formula is proposed for all leptons $\nu_1, 
\nu_2 , \nu_3$ and $e^-,\,\mu^-,\,\tau^- $ as well as all quarks $u,\,c,\,t$ and $d,\,s,\,b$ of three 
generations, parametrized by three free constants $ \mu, \varepsilon, \xi$ assuming 
four different triplets of values. Four such triplets of parameter values are determined or 
estimated from the present data. Mass spectra in the four cases are related to each other by 
{\it shifting} the triplet of parameters $ \mu, \varepsilon, \xi$. For charged leptons 
$\xi \simeq 0$ (but probably $\xi \neq 0$). If for them $\xi $ is put to be exactly 0, then $m_\tau = 1776.80$ MeV is {\it predicted} after the input of experimental $m_e$ and $m_\mu$ (the central 
value of experimental $m_\tau = 1776.99^{+0.29}_{-0.26}$ MeV corresponds to  $ \xi =1.8\times 10^{-3}\neq 0$). For neutrinos $1/\xi \simeq 0$ (but $1/\xi \neq 0$ in the case of normal hierarchy 
$m^2_{\nu_1} \ll m^2_{\nu_2} \ll m^2_{\nu_3}$). If for neutrinos $1/\xi $ is conjectured to be exactly 
0, then $(m_{\nu_1}, m_{\nu_2}, m_{\nu_3}) \sim (1.5, 1.2, 5.1)\times 10^{-2}$ eV are {\it predicted} 
after the input of experimental estimates $|m^2_{\nu_2} - m^2_{\nu_1}| \sim 8.0\times 
10^{-5}\;{\rm eV}^2$ and $|m^2_{\nu_3} - m^2_{\nu_2}| \sim 2.4\times 10^{-3}\;{\rm eV}^2$. 
Thus, the mass ordering of neutrino states 1 and 2 is then {\it inverted}, while the 
position of state 3 is {\it normal}.

\vspace{0.5cm}

\ni PACS numbers: 12.15.Ff , 14.60.Pq  .

\vspace{0.6cm}

\ni January 2006  

\vfill\eject

~~~
\pagestyle {plain}

\setcounter{page}{1}

\vspace{0.3cm}

\ni {\bf 1. Introduction}

\vspace{0.3cm}

Some time ago we have found an efficient two-parameter mass formula for charged leptons, predicting reasonably the mass $m_\tau$ from the input of experimental masses $m_e$ and $m_\mu$ [1]. Then, we have extended this formula to up and down quarks, introducing necessarily a third parameter [2]. Recently, we have considered a few versions of two- and three-parameter mass formula for active neutrinos [3]. In the present paper, we propose a specific {\it universal} shape of empirical mass formula for all leptons $\nu_1, \nu_2, \nu_3$ and $e^-,\,\mu^-,\,\tau^- $ as well as all quarks $u,\,c,\,t$ and $d,\,s,\,b$ of three generations, parametrized by three free constants assuming  four different sets of three values. This mass formula reads :

\begin{equation}
m_i  =  \mu\, \rho_i \left(N^2_i + \frac{\varepsilon -1}{N^2_i} - \xi\right) \;\;(i = 1,2,3)\;,
\end{equation}

\ni where the numbers

\begin{equation}
N_1 = 1 \;,\; N_2 = 3 \;,\; N_3 = 5
\end{equation}

\ni and

\begin{equation}
\rho_1 = \frac{1}{29} \;,\; \rho_2 = \frac{4}{29} \;,\; \rho_3 = \frac{24}{29}
\end{equation}

\ni ($\sum_i \rho_i = 1$) are {\it fixed} elements in all four cases, while $ \mu , \varepsilon , \xi $ are three free parameters that take four different sets of three parameter values (the normalized fractions $\rho_i$ may be called "generation-weighting factors"). Here,

\vspace{0.2cm}

\begin{equation}
(m_1, m_2, m_3) = \left\{ \begin{array}{rl} (m_{\nu_1}, m_{\nu_2}, m_{\nu_3}) & {\rm for \; active \; neutrinos}, \\ (m_e\,, \, m_\mu\,, \, m_\tau\,) & {\rm for \; charged \; leptons}, \\ (m_u\,,\, m_c\,, \, m_t\;) & {\rm for \; up \; quarks}, \\ (m_d\,, \, m_s\,, \, m_b\;) & {\rm for \; down \; quarks} 
\end{array}\right.
\end{equation}

\vspace{0.3cm}

\ni are experimental masses (as we know them). Strictly speaking, $m_{\nu_i}$ in Eq. (1) are neutrino Dirac masses $m^{(D)}_{\nu_i}$, subject to recalculation into active-neutrino masses $m_{\nu_i}$. The active mass neutrinos $\nu_i \;(i=1,2,3)$ are related to the active weak-interaction neutrinos $\nu_\alpha \;(\alpha = e^-,\mu^-, \tau^-)$ through the familiar unitary transformation $\nu_\alpha = \sum_i U_{\alpha i} \nu_i$.

From Eq. (1), rewritten in the explicit form

\begin{eqnarray}
m_1 & = & \frac{\mu}{29} (\varepsilon - \xi) \,, \\
m_2 & = & \frac{\mu}{29} \frac{4}{9} (80 +\varepsilon - 9\,\xi) \,, \\
m_3 & = & \frac{\mu}{29} \frac{24}{25} (624 + \varepsilon - 25\,\xi) \,, 
\end{eqnarray}

\ni we can evaluate the parameters:

\vspace{-0.2cm}

\begin{eqnarray}
\mu & = & \frac{29\cdot 25}{1536\cdot 6} \left[m_3 - \frac{6}{25}(27 m_2 - 8m_1)\right] \,, \\
\varepsilon & = & 10 \frac{m_3 - \frac{6}{125}(351 m_2 - 904 m_1)}{m_3 - \frac{6}{25}(27 m_2 - 8m_1)} \,, \\
\xi & = & 10 \frac{m_3 - \frac{6}{125}(351 m_2 - 136 m_1)}{m_3 - \frac{6}{25}(27 m_2 - 8m_1)}  \,, 
\end{eqnarray}

\ni and find also the mass sum rule:

\vspace{-0.2cm}

\begin{equation}
m_3 = \frac{1}{1-\xi/10}\;\frac{6}{125}\left[351(1 - \xi/26)m_2 - 136 (1 - \xi/34) m_1\right] \,. 
\end{equation}

\ni Note also the relations 

\begin{equation}
\mu = \frac{29 m_1}{\varepsilon - \xi} =  \frac{1}{1- \xi/10}\; \frac{29 (9m_2 - 4 m_1)}{320}
\end{equation}

\ni and

\vspace{-0.2cm}

\begin{equation}
\varepsilon - \xi = (1- \xi/10) \;\frac{320 m_1}{9 m_2 - 4 m_1}
\end{equation}

\ni following from the formula (1).

Since the shape $m_i = F_i(\mu,\varepsilon,\xi)\;(i = 1,2,3)$ of mass formula (1) is {\it the same} in four cases of fundamental fermions, the four mass spectra of them are related to each other by {\it shifting} the set of three parameters $\mu, \varepsilon, \xi$. Three parameters $\mu, \varepsilon, \xi $, assuming four different sets of three parameter values, determine four mass spectra of fundamental fermions. Then, the mass formula (1) gives $ m_{f_i} = F_i(\mu^{(f)}, \varepsilon^{(f)}, \xi^{(f)})\;(i=1,2,3)$, where $f = \nu, l, u , d$ labels four kinds of fundamental fermions: neutrinos, charged leptons, up quarks and down quarks, respectively (the function $F_i(\mu, \varepsilon, \xi)$ is {\it universal i.e.}, independent of the label $f$). Strictly speaking, in the case of neutrinos, the mass formula (1) gives directly three neutrino Dirac masses $m^{(D)}_{\nu_i}\;(i=1,2,3)$ that generically ought to be recalculated afterwards into three physical active-neutrino masses $m_{\nu_i}\;(i=1,2,3)$ through the seesaw mechanism (or another analogical procedure).

The mass formula (1) may lead to some specific {\it predictions} for $m_{f_1}, m_{f_2}, m_{f_3}$ (some specific relations for them), if {\it not} all three $\mu^{(f)}, \varepsilon^{(f)}, \xi^{(f)}$ for a particular $f$ are really free parameters, for instance, if one of them happens to be fixed ({\it e.g.}, if $\xi^{(l)} = 0$ or if $1/\xi^{(\nu)} = 0$), while two others remain free parameters determined by the input of two of $m_{f_1}, m_{f_2}, m_{f_3}$ and so, predict the third of these masses through the formula (1). 

\vspace{0.3cm}

\ni {\bf 2. Charged leptons}

\vspace{0.3cm}

In this case, the experimental masses are [4]

\vspace{-0.2cm}

\begin{equation}
m_{e} = 0.5109989\;{\rm MeV} \;,\; m_{\mu} = 105.65837\;{\rm MeV} \;,\; m_{\tau} = 1776.99^{+0.29}_{-0.26}\;{\rm MeV} \;.
\end{equation}

\ni Thus, from Eqs. (8) and (9) we find (with the central value of $m_\tau$)

\begin{equation}
\mu = 86.0076\;{\rm MeV} \;,\;\varepsilon = 0.174069 \;,
\end{equation}

\ni and from Eq. (12)

\vspace{-0.2cm}

\begin{equation}
\varepsilon - \xi = \frac{29 m_e}{\mu} = 0.172298 \;.
\end{equation}

\ni Hence, 

\vspace{-0.2cm}

\begin{equation}
\xi \equiv \varepsilon - (\varepsilon - \xi) = 1.771\times 10^{-3} =1.8\times 10^{-3} \,.
\end{equation}

\ni The same value of $\xi$ follows from Eq. (10). Notice that $\xi$ for charged leptons is very small with respect to the terms $N^2_i + (\varepsilon -1)/N^2_i $ in Eq. (1).

If we put for charged leptons exactly $\xi = 0$, we evaluate from Eqs. (12) and (13)

\begin{equation}
\mu = \frac{29 (9m_\mu - 4 m_e)}{320} = 85.9924\;{\rm MeV}
\end{equation}

\ni and

\vspace{-0.2cm}

\begin{equation}
\varepsilon = \frac{320 m_e}{9 m_\mu - 4 m_e} = 0.172329,
\end{equation}

\ni respectively, and {\it predict} from Eq. (11) [1]

\begin{equation} 
m_\tau = \frac{6}{125} (351m_\mu -  136 m_e) = 1776.80\;{\rm MeV} 
\end{equation}

\ni in a very good agreement with experimental $m_\tau $ given in Eq. (14). In calculating the values (18) and (19) for $\mu$ and $\varepsilon$ we use as an input only the experimental $m_e$ and $m_\mu$. Of course, we reproduce exactly all three values (14) of masses $m_e, m_\mu$ and $m_\tau $ (its central value), when we make use of three exact values (15) and (17) of parameters $\mu, \varepsilon$ and $\xi$.

\vspace{0.3cm}

\ni {\bf 3. Up and down quarks}

\vspace{0.3cm}

In the case of up and down quarks, the medium experimental mass values are [4]

\vspace{-0.2cm}

\begin{equation} 
m_u \sim 2.8 \;{\rm MeV}\;,\;  m_c \sim 1.3 \;{\rm GeV}\;,\;  m_t \sim 174 \;{\rm GeV} 
\end{equation}

\ni and

\vspace{-0.2cm}

\begin{equation} 
m_d \sim 6 \;{\rm MeV}\;,\;  m_s \sim 110 \;{\rm MeV}\;,\;  m_b \sim 4.3\;{\rm GeV} \,,
\end{equation}

\ni respectively. Thus, using Eqs. (8), (9) and (10) we obtain

\vspace{-0.2cm}

\begin{equation} 
\mu \sim 13 \;{\rm GeV}\;,\; \varepsilon \sim 9.2 \;,\; \xi \sim 9.2
\end{equation}

\ni and

\vspace{-0.2cm}

\begin{equation} 
\mu \sim 280 \;{\rm MeV}\;,\; \varepsilon \sim 7.5 \;,\; \xi \sim 6.9 \,,
\end{equation}

\ni respectively. More precisely,

\vspace{-0.2cm}

\begin{equation} 
\varepsilon - \xi = \frac{29 m_u}{\mu} \sim 0.0062 \;,\; \varepsilon/\xi - 1 \sim 6.7\times 10^{-4}
\end{equation}

\ni and

\vspace{-0.2cm}

\begin{equation} 
\varepsilon - \xi = \frac{29 m_d}{\mu} \sim 0.61 \;,\; \varepsilon/\xi - 1 \sim 8.8\times 10^{-2} \,,
\end{equation}

\ni respectively. Of course, we can reproduce all quark masses (21) and (22), when we use the values (23) and (24) of parameters $\mu, \varepsilon$, $\xi $ (and also Eqs. (25) and (26)).

We can see that for up and down quarks $\xi \simeq \varepsilon$ (especially for up quarks). If we put for them exactly $\xi = \varepsilon$, we {\it predict} from Eq. (5) that $m_u = 0$ and $m_d = 0$, and then evaluate from Eqs. (8) and (10) that $\mu \sim 13$ GeV and $\varepsilon = \xi \sim 9.2$ for up quarks and $\mu \sim 280$ MeV and $\varepsilon = \xi \sim 6.8$ for down quarks.

\vspace{0.3cm}

\ni {\bf 4. Neutrinos}

\vspace{0.3cm}

The situation for neutrinos may be different than for three other kinds of fundamental fermions since, being electrically neutral, they may be Majorana fermions, in contrast to the others which are Dirac fermions. Denote by $\nu_i \equiv \nu_{iL}$ and $N_i \equiv \nu_{iR}\;(i=1,2,3)$ the three active (lefthanded) and three sterile (righthanded) mass neutrinos, and by $m_{\nu_i}$ and $ m_{N_i}$ their respective masses, being eigenstates of the corresponding $3\times 3$ mass matrices $M^{(\nu)}$ and $M^{(N)}$ (in the flavor basis). Of course, the righthanded-neutrino mass states $N_i$ must not be confused with the numbers $N_i = 1,3,5$. 

Assume that the seesaw mechanism works and that $M^{(N)}$ commutes with the neutrino Dirac $3\times 3$ mass matrix $M^{(D)}$ (in the flavor basis), giving the neutrino Dirac masses $m_{\nu_i}^{(D)}\;(i=1,2,3)$ as its eigenstates. Then, we can write

\vspace{-0.2cm}

\begin{equation} 
m_{\nu_i} = - \frac{ m_{\nu_i}^{(D)\,2}}{m_{N_i}} \,,
\end{equation}

\ni as a consequence of the popular seesaw relation

\vspace{-0.2cm}

\begin{equation} 
M^{(\nu)} = - M^{(D)\,T} \frac{1}{M^{(N)}} M^{(D)} 
\end{equation}

\ni with $M^{(D)\,\dagger} = M^{(D)}$ and $M^{(N)\,T} = M^{(N)}$. Now, make (for simplicity) the conjecture that the assumed commutation $[M^{(N)}, M^{(D)}] = 0$ is realized (trivially) by the matrix proportionality $M^{(N)} = \zeta\, M^{(D)}$ with a very large parameter $\zeta >0$. This implies the eigenvalue proportionality 

\vspace{-0.2cm}

\begin{equation} 
m_{N_i} =  \zeta m_{\nu_i}^{(D)} \,.
\end{equation}

\ni Then, from Eqs. (27) and (29) we obtain 

\vspace{-0.2cm}

\begin{equation} 
m_{\nu_i} = - (1/\zeta) m_{\nu_i}^{(D)} = - (1/\zeta)^2 m_{N_i} \,.
\end{equation}

\ni Hence, the commutation $[M^{(\nu)}, M^{(N)}] = 0$. If $m_{\nu_i} > 0$, then it follows that $m_{N_i} < 0$ from Eq. (27) and $m^{(D)}_{\nu_i} < 0$ from Eq. (30) (masses for relativistic spin-1/2 fields may be negative, since only masses squared can be physical).

Now, conjecture for neutrino Dirac masses $m^{(D)}_{\nu_i}$ our mass formula (1):

\vspace{-0.2cm}

\begin{equation}
m^{(D)}_{\nu_i}  = \mu \, \rho_i \left(N^2_i + \frac{\varepsilon -1}{N^2_i} - \xi\right) \;\;(i = 1,2,3)\;.
\end{equation}

\ni Then, making use of the seesaw relation (30) (valid in the case of the proportionality (29)), we obtain for active neutrinos $\nu_i$ the following mass formula:

\vspace{-0.2cm}

\begin{equation}
m_{\nu_i}  = (\mu \,\xi/\zeta) \rho_i \left[1- (1/\xi) \left(N^2_i + \frac{\varepsilon -1}{N^2_i}\right)\right] \;\;(i = 1,2,3)\;.
\end{equation}

Hence, rewriting this formula in the explicit form 

\vspace{-0.2cm}

\begin{eqnarray}
m_{\nu_1} & = & \frac{\mu \,\xi/\zeta}{29} (1 - \varepsilon/\xi) \,, \\
m_{\nu_2} & = & \frac{\mu \,\xi/\zeta}{29} 4\left[1 -\frac{1}{9}(1/\xi)(80 + \varepsilon)\right] \,, \\
m_{\nu_3} & = & \frac{\mu \,\xi/\zeta}{29} 24\left[ 1 -\frac{1}{25}(1/\xi) (624 + \varepsilon)\right] \,, 
\end{eqnarray}

\ni we can evaluate the parameters $\mu \,\xi/\zeta\,,\,\varepsilon\,,\,1/\xi$ in terms of the 
active-neutrino masses $m_{\nu_1}, m_{\nu_2},m_{\nu_3}$:

\vspace{-0.2cm}

\begin{eqnarray}
\mu \xi/\zeta & = & -\frac{29\cdot 125}{1536\cdot 3} \left[m_{\nu_3} - \frac{6}{125}(351 m_{\nu_2} - 136m_{\nu_1})\right] \,, \\
\varepsilon & = & 10\, \frac{m_{\nu_3} - \frac{6}{125}(351 m_{\nu_2} - 904 m_{\nu_1}) }{m_{\nu_3} - \frac{6}{25}(27 m_{\nu_2} - 8m_{\nu_1})} \,, \\
1/\xi & = & \frac{1}{10}\; \frac{m_{\nu_3} - \frac{6}{25}(27 m_{\nu_2} - 8m_{\nu_1})}{m_{\nu_3} - \frac{6}{125}(351 m_{\nu_2} - 136 m_{\nu_1})}  \,, 
\end{eqnarray}

\ni and find also the mass sum rule: 

\vspace{-0.2cm}

\begin{equation}
m_{\nu_3} = \frac{1}{1- 10/\xi}\;\frac{6}{25}\left[27(1 - 26/\xi)m_{\nu_2} - 8 (1 - 34/\xi) m_{\nu_1}\right] \,. 
\end{equation}

\ni Note also the relations 

\vspace{-0.4cm}

\begin{equation}
\mu\, \xi/\zeta = \frac{29 m_{\nu_1}}{1 -\varepsilon/\xi} =  \frac{1}{1- 10/\xi}\; \frac{29}{32} (9m_{\nu_2} - 4 m_{\nu_1})
\end{equation}

\ni and

\vspace{-0.4cm}

\begin{equation}
1 -\varepsilon/\xi = (1- 10/\xi) \;\frac{32 m_{\nu_1}}{9 m_{\nu_2} - 4 m_{\nu_1}}
\end{equation}

\ni following from the mass formula (32). 

Unfortunately, in spite of the enormous progress in neutrino physics, data on neutrino masses are still unsatisfactory. In fact, only the mass-squred differences are reasonably well estimated [5]:

\vspace{-0.4cm}

\begin{equation}
|\Delta m^2_{21}| \equiv |m^2_{\nu_2} - m^2_{\nu_1}| \sim 8.0\times 10^{-5}\; {\rm eV}^2 \;,\; 
|\Delta m^2_{32}| \equiv |m^2_{\nu_3} - m^2_{\nu_2}| \sim 2.4\times 10^{-3}\; {\rm eV}^2 \;,
\end{equation}

\ni giving in the case of normal hierarchy $m^2_{\nu_1} \ll m^2_{\nu_2} \ll m^2_{\nu_3}$ the estimations

\vspace{-0.2cm}

\begin{eqnarray}
m_{\nu_2} & \equiv & \sqrt{\Delta m^2_{21} + m_{\nu_1}} \simeq \sqrt{\Delta m^2_{21}}  \sim 8.9\times 10^{-3}\;{\rm eV} \;, \nonumber \\ m_{\nu_3} & \equiv & \sqrt{\Delta m^2_{32} + \Delta m^2_{21} + m_{\nu_1}} \simeq \sqrt{\Delta m^2_{32} + \Delta m^2_{21}} \sim 5.0\times 10^{-2}\;{\rm eV} \;,
\end{eqnarray}

\ni when $m^2_{\nu_1}$ can be neglected (it cannot be neglected {\it e.g.} for $m_{\nu_1} \sim 1\times 10^{-3}$ eV, where $m_{\nu_2} \sim 9.0\times 10^{-3}$ eV and $m_{\nu_3} \sim 5.0\times 10^{-2}$ eV). Their ratios are

\vspace{-0.2cm}

\begin{equation}
\frac{\Delta m^2_{32}}{\Delta m^2_{21}} \sim 30 \;,\; \frac{m_{\nu_3}}{m_{\nu_2}} \sim \sqrt{30} = 5.5 \;.
\end{equation}

Since in the case of normal hierarchy $m^2_{\nu_1} \ll m^2_{\nu_2} \ll m^2_{\nu_3}$ the mass $m_{\nu_1}$ is very small, consider the following range of its possible value

\vspace{-0.2cm}

\begin{equation}
m_{\nu_1} \sim(0 \;{\rm to} \;1)\times 10^{-2}\;{\rm eV} \;.
\end{equation}

\ni Then, $m_{\nu_2} \sim (8.9\; {\rm to}\; 9.0)\times 10^{-3}$ eV and $m_{\nu_3} \sim 5.0\times 10^{-2}$ eV ({\it cf.} Eqs. (42)). In this situation, we can evaluate from Eq. (36)

\vspace{-0.2cm}

\begin{equation}
\mu \xi/\zeta \sim (7.9\;{\rm to} \;7.5)\times 10^{-2}\;{\rm eV} \;,
\end{equation}

\ni from Eq. (37)

\vspace{-0.2cm}

\begin{equation}
\varepsilon \sim (120\;{\rm to} \;89) 
\end{equation}

\ni and from Eq. (38)

\vspace{-0.2cm}

\begin{equation}
1/\xi \sim (8.1\;{\rm to} \;6.9)\times 10^{-3} \;,\; \varepsilon/\xi \sim (1\;{\rm to} \;0.61) \;.
\end{equation}

\ni Of course, we can reproduce all active-neutrino masses $m_{\nu_1}, m_{\nu_2}, m_{\nu_3}$, if we know adequate values of the parameters $\mu, \varepsilon, \xi$.

It is interesting to notice that, in the case of normal hierarchy $m^2_{\nu_1} \ll m^2_{\nu_2} \ll m^2_{\nu_3}$, the parameter $1/\xi $ for neutrinos is very small, though not 0 ({\it cf.} Eqs. (48)). In contrast, for charged leptons, the parameter $\xi$ is very small, but not 0 for the central value of experimental $m_\tau$ ({\it cf.} Eq. (17)). This suggests the existence of  a kind of {\it complementarity} between neutrinos and charged leptons.

It seems worthwhile to investigate for neutrinos the strict limit of $1/\xi \rightarrow 0$ that implies a new mass sum rule (49) below. Similarly, in the case of charged leptons, the strict limit of $\xi \rightarrow 0$ leads to the mass sum rule (20) which is consistent with their experimental masses within the uncertainty limits of $m_\tau $. To this end observe that, in the limit of $1/\xi \rightarrow 0$, Eq. (39) gives the simplified sum rule:

\vspace{-0.2cm}

\begin{equation}
m_{\nu_3} = \frac{6}{25}\left(27m_{\nu_2} - 8 m_{\nu_1}\right) \,,
\end{equation}

\ni implying the equality

\vspace{-0.4cm}

\begin{equation}
\left[\frac{6}{25}(27m_{\nu_2} - 8m_{\nu_1})\right]^2 - m^2_{\nu_2} = {\Delta m^2_{32}} \equiv \frac{\Delta m^2_{32}}{\Delta m^2_{21}}\;(m^2_{\nu_2} - m^2_{\nu_1}) \,. 
\end{equation}

\ni Denoting

\vspace{-0.4cm}

\begin{equation}
r \equiv \frac{ m_{\nu_2}}{ m_{\nu_1}}\;,\; \lambda \equiv \frac{\Delta m^2_{32}}{\Delta m^2_{21}} \sim \pm 30
\end{equation}

\ni and dividing Eq. (50) by $m^2_{\nu_1}$, we find the following quadratic equation for $r$: 

\begin{equation}
\left(\frac{6}{25}\right)^2 (27r - 8)^2 - (1+\lambda) r^2 + \lambda = 0 \,. 
\end{equation}

\ni With $\lambda \sim 30 > 0$, this equation gets two complex sulutions for $r$ that cannot be physical. Thus, in the strict limit of $1/\xi \rightarrow 0$ the neutrino mass orderings $m^2_{\nu_1} < m^2_{\nu_2} < m^2_{\nu_3}$ and $m^2_{\nu_3} < m^2_{\nu_2} < m^2_{\nu_1}$ leading to $\lambda>0$ are both excluded. On the contrary, with $\lambda \sim -30 < 0$, there appear two real solutions for $r$:

\begin{equation}
r \sim \left\{\begin{array}{rr} -0.46 & <0 \\ 0.81 & >0 \end{array}\right. \;,
\end{equation}

\ni corresponding to

\vspace{-0.2cm}

\begin{equation}
m_{\nu_2} \equiv r \,m_{\nu_1} \left\{\begin{array}{rr} -0.46\, m_{\nu_1}  & <0 \\ 0.81 \,m_{\nu_1}  & >0 \end{array}\right. \;,
\end{equation}

\ni where we choose $m_{\nu_1} >0$ (here and below, only two decimals are significant as only two are such in $\lambda $). The requirement that all three masses $m_{\nu_i}$ of active-neutrino triplet $\nu_1, \nu_2, \nu_3 $ should have the same sign, excludes the first solution (54). Then, only the second solution (54) remains as physical. 

For such a unique solution it follows that

\begin{equation}
\Delta m^2_{21}\equiv (r^2 -1)m^2_{\nu_1} \sim -0.35\, m^2_{\nu_1} < 0
\end{equation}

\ni and so, $\Delta m^2_{21} \sim - 8.0\times 10^{-5}\;{\rm eV}^2 < 0$, while $\Delta m^2_{32} \sim 2.4\times 10^{-3}\;{\rm eV}^2 > 0$ as $\lambda <0$ in this case. Thus, from Eq. (55)

\begin{eqnarray}
m^2_{\nu_1} & \equiv & \frac{\Delta m^2_{21} }{r^2-1} \sim 2.3\times 10^{-4}\;{\rm eV}^2   \,, \\
m^2_{\nu_2} & \equiv & \Delta m^2_{21} +m^2_{\nu_1} \sim 1.5\times 10^{-4}\;{\rm eV}^2 
\end{eqnarray}

\ni and

\vspace{-0.2cm}

\begin{equation}
m^2_{\nu_3} \equiv \Delta m^2_{32} +m^2_{\nu_2} \sim 2.6\times 10^{-3}\; {\rm eV}^2 \,. 
\end{equation}

\ni This gives $\Delta m^2_{31} \equiv m^2_{\nu_3} - m^2_{\nu_1} \sim 2.3\times 10^{-3}\; {\rm eV}^2 $ and

\begin{equation}
m^2_{\nu_2} : m^2_{\nu_1} : m^2_{\nu_3} \sim 0.65 : 1 : 11\,.
\end{equation}

\ni We can see that

\vspace{-0.2cm}

\begin{equation}
m^2_{\nu_2} < m^2_{\nu_1} < m^2_{\nu_3} 
\end{equation}

\ni {\it i.e.}, the mass ordering of neutrino states 1 and 2 is {\it inverted}, while the position of neutrino state 3 is {\it normal}. The hierarchy is here more moderate than in the case of very small $m_{\nu_1}$ (Eq. (45)). From Eqs. (56), (57) and (58) we {\it predict} that

\vspace{-0.2cm}

\begin{equation}
m_{\nu_1} \sim 1.5\times 10^{-2}\;{\rm eV}\;,\;m_{\nu_2} \sim 1.2\times 10^{-2}\;{\rm eV}\;,\; m_{\nu_3} \sim 5.1\times 10^{-2}\;{\rm eV}\;,
\end{equation}

\ni implying the proportion

\vspace{-0.2cm}

\begin{equation}
m_{\nu_2} : m_{\nu_1} : m_{\nu_3} \sim 0.81 : 1 : 3.3\,.
\end{equation}

\ni It is easy to check that these masses of active neutrinos really satisfy the mass sum rule (49) valid in the strict limit of $1/\xi \rightarrow 0$. Recall that the ordering of indices $i =1,2,3$ is established by the form of neutrino mixing matrix $U = (U_{\alpha i})$ transforming active 
weak-interaction neutrinos $\nu_\alpha \;(\alpha = e,  \mu, \tau)$ into active mass neutrinos $\nu_i \;(i = 1,2,3)$.

With the values (61) of $m_{\nu_1}$ and $m_{\nu_2}$ we can evaluate $\mu \xi/\zeta $ and $\varepsilon/\xi$ in the strict limit of $1/\xi \rightarrow 0$, applying Eqs. (40) and (41) considered in this limit:

\begin{equation}
\mu \xi/\zeta \rightarrow \frac{29}{32}(9 m_{\nu_2} - 4 m_{\nu_1}) \sim 4.5\times 10^{-2}\;{\rm eV}
\end{equation}

\ni and

\vspace{-0.2cm}

\begin{equation}
\varepsilon/\xi \rightarrow 1 - \frac{32 m_{\nu_1}}{9 m_{\nu_2} - 4 m_{\nu_1}} \sim -8.8 < 0 \,.
\end{equation}

\ni Thus, $\mu \rightarrow +0$ and $1/\varepsilon \rightarrow -0$ when $1/\xi \rightarrow +0$, but only the constants $\mu \xi/\zeta $ and  $\varepsilon/\xi $ as well as $1/\xi $ appear in the mass formula (32) which, in the limit of $1/\xi \rightarrow 0$, takes the form

\begin{equation}
m_{\nu_i} = \mu' \rho_i \left(1 + \frac{\varepsilon'}{N^2_i}\right) \,,
\end{equation}

\ni where $\mu \xi/\zeta \rightarrow \mu' \sim 4.5\times 10^{-2}\;{\rm eV}$ and $-\varepsilon/\xi \rightarrow \varepsilon' \sim 8.8 > 0 $.

Finally, not passing with $1/\xi$ strictly to 0, consider $1/\xi$ smaller than the values in Eq. (48) corresponding to the range $m_{\nu_1} \sim (0\;{\rm to}\; 1)\times 10^{-3}$ eV (Eq. (45)), where still  $m^2_{\nu_1} \ll m^2_{\nu_2} \ll m^2_{\nu_3}$. For instance, put $1/\xi \sim  1.8\times 10^{-3}$ eV which is the value of neutrino $1/\xi \equiv 1/\xi^{(\nu)}$ related to the charged-lepton $\xi^{(l)}$ (Eq. (17)) through the simplest form of complementarity: $\xi^{(\nu)} \xi^{(l)} \sim 1$ or $1/\xi^{(\nu)} \sim \xi^{(l)} = 1.8\times 10^{-3}$. In the case of arbitrary $1/\xi$, we derive for the ratio $r \equiv m_{\nu_2}/m_{\nu_1}$ the following generalized form of Eq. (52) (the latter being valid in the limit $1/\xi \rightarrow 0$):

\begin{equation}
\frac{1}{(1-10/\xi)^2}\,\left(\frac{6}{25}\right)^2 [27(1-26/\xi)r - 8(1- 34/\xi)]^2 - (1+\lambda) r^2 + \lambda = 0 \,,
\end{equation}

\ni where $|\lambda| \sim 30$. Now, consider $1/\xi \sim 1.8\times 10^{-3}$. For $\lambda \sim 30$, both solutions to Eq. (66) turn out to be complex. For $\lambda \sim -30$, both solutions are real, but only one positive:

\begin{equation}
r \sim 0.81> 0\,,
\end{equation}

\ni corresponding to $m_{\nu_2} \sim 0.81 m_{\nu_1}$ (a difference with the limiting value (53) appears at the level of further decimals). Then, with $|\Delta m^2_{21}| \sim 8.0\times 10^{-5}\; {\rm eV}^2$ and $|\Delta m^2_{32}| \sim 2.4\times 10^{-3}\; {\rm eV}^2$ we obtain 

\begin{equation}
m^2_{\nu_1} \sim 2.4\times 10^{-4}\; {\rm eV}^2 \,,\, m^2_{\nu_2} \sim 1.6\times 10^{-4}\; {\rm eV}^2 \,,\, m^2_{\nu_3} \sim 2.6\times 10^{-4}\; {\rm eV}^2 
\end{equation}

\ni and

\begin{equation}
m_{\nu_1} \sim 1.5\times 10^{-2}\; {\rm eV} \,,\, m_{\nu_2} \sim 1.3\times 10^{-2}\; {\rm eV} \,,\, m_{\nu_3} \sim 5.1\times 10^{-2}\; {\rm eV} \,.
\end{equation}

\ni Thus, the mass ordering of 1 and 2 neutrino states is {\it inverted} ($\Delta m^2_{21} \sim -8.0 \times 10^{-5}\; {\rm eV}^2$), but the position of 3 neutrino state is {\it normal} ($\Delta m^2_{32} \sim 2.4\times 10^{-3}\; {\rm eV}^2$ as $\lambda < 0$).

In the case of $1/\xi \sim 1.8\times 10^{-3}$, using the values (69), we can also evaluate

\begin{equation}
\mu \xi/ \zeta = \frac{29}{32}(9 m_{\nu_2} - 4m_{\nu_1}) \sim 4.6\times 10^{-2}\;{\rm eV}\; ,\;\mu/\zeta \sim 8.3\times 10^{-5}\;{\rm eV}\; 
\end{equation}

\ni and

\begin{equation}
\varepsilon/\xi = 1 -\frac{29 m_{\nu_1}}{\mu \xi/\zeta} \sim -8.7 \;,\; \varepsilon \sim -4800 \,.
\end{equation}

\ni So, we can see that, in this case, the results (67) -- (71) at the level of two decimals are practically identical to those in the strict limit of $1/\xi \rightarrow 0$  (except for $\mu/\zeta $ and $\varepsilon $ which could not be obtained with $\xi \rightarrow \infty $).

\vspace{0.3cm}

\ni {\bf 5. Conclusions}

\vspace{0.3cm}

In this paper, we have proposed a specific {\it universal} shape (1) of empirical mass formula for all fundamental fermions: leptons $\nu_1, \nu_2 , \nu_3$ and $e^-,\,\mu^-,\,\tau^- $ as well as quarks $u,\,c,\,t$ and $d,\,s,\,b$ of three generations, parametrized by four different sets of three  free constants $\mu, \varepsilon, \xi $. Mass spectra in the four cases are related to each other by {\it shifting} the set of three parameters $\mu, \varepsilon, \xi $. The parameter $\mu$ plays the role of "radius"\, in the three-dimentional mass space of $m_1, m_2, m_3$, while $\varepsilon $ and $\xi $ are connected with "spherical angles"\, in this space. Strictly speaking, for active neutrinos the mass formula holds in the form (32) related (in the seesaw mechanism) to the primary mass formula (1), when the latter is valid for neutrino Dirac masses and when the matrix proportionality $M^{(N)} = \zeta\, M^{(D)}$ with a very large parameter $\zeta >0$ is assumed (for simplicity). In the mass formula (32) the primary parameters $\mu $ and $\varepsilon $ are replaced by $\mu \xi/\zeta$ and $\varepsilon/\xi$. Note that the seesaw $\zeta$ parameter
is equal to

\begin{equation}
\zeta \equiv \frac{\mu^{(l)}}{\mu^{(\nu)}/\zeta}\,\frac{\mu^{(\nu)}}{\mu^{(l)}} \sim \frac{86\times 10^6}{(6.4 \;{\rm to}\; 5.2)\times 10^{-4}}\,\frac{\mu^{(\nu)}}{\mu^{(l)}} = (1.3 \;{\rm to}\; 1.7)\times 10^{11}\, \frac{\mu^{(\nu)}}{\mu^{(l)}} \,,
\end{equation}

\ni where $\mu^{(l)} = 86.0076$ MeV and $\mu^{(\nu)}/\zeta \sim (6.4 \;{\rm to}\;5.2)\times 10^{-4}$ eV for charged leptons and neutrinos, respectively, as can be seen from our listing below. If $\mu^{(\nu)} \simeq \mu^{(l)}$, then $\zeta = O(10^{11})$. If rather  $\mu^{(\nu)} : \mu^{(l)} \simeq \mu^{(u)} : \mu^{(d)} \sim 46$ (Eqs. (23) and (24)), then $\zeta = O(10^{13})$.

From the mass formula (1) we have evaluated or estimated the following parameter values: 

$$
\begin{array}{rccc} \!\!\!\!{\rm for}\, e^-\!\!,\mu^-\!,\tau^-\!:\!\!\! & \!\!\mu = 86.0076 \;{\rm MeV}\,, & \!\!\varepsilon = 0.174069 \,, & \!\!\xi = 1.8\times 10^{-3}\,, \\ \!\!{\rm when}\;\; & \!\!m_e = 0.510999 \,{\rm MeV}  , & \!\!m_\mu = 105.658 \,{\rm MeV}, & \!\!m_\tau = 1776.99 \;{\rm MeV} \,, \\ \\
\!\!\!\!{\rm for}\;\, u\,, \,c\,, \,t\, :\!\! & \!\!\mu \sim 13 \;{\rm GeV}\,, & \!\!\varepsilon \sim 9.2\,, & \!\!\xi \sim 9.2\,, \\ \!\!{\rm when}\;\; & \!\!m_u \sim 2.8 \;{\rm MeV}\,, & \!\!m_c \sim 1.3 \;{\rm GeV} , & \!\!m_t \sim 174 \,{\rm GeV}\,, \\ \\ 
\!\!\!\!\!{\rm for}\;\, d\,, \,s\,, \,b\, :\!\! & \!\!\mu \sim 280 \;{\rm MeV}\,, & \!\!\varepsilon \sim 7.5\,, & \!\!\xi \sim 6.9\,, \\ \!\!{\rm when}\;\; & \!\!m_d \sim 6 \;{\rm MeV}\,, & \!\!m_s \sim 110 \;{\rm MeV}\,, & \!\!m_b \sim 4.3 \,{\rm GeV}\,, \\ \\ 
\!\!\!\!\!{\rm for}\;\nu_1,\nu_2,\nu_3 :\!\! & \!\!\mu \xi/\zeta \!\!\sim\!(7.9\,{\rm to}\,7.5)\!\!\times\! 10^{-2}\,{\rm eV}, & \!\!\varepsilon/\xi \!\sim \!(1\,{\rm to}\,0.61)\,, & \!\!1/\xi \!\!\sim \!(8.1\,{\rm to}\,6.9)\!\!\times\! \!10^{-3},\\ 
\!\!\left[{\rm and}\;\;\right. & \!\!\!\!\mu/\zeta \!\!\sim \!(6.4\,{\rm to}\,5.2)\!\!\times\!\! 10^{-4}\,{\rm eV},\!\! & \!\!\varepsilon \!\sim \!(120\,{\rm to}\,89), & \!\!\!\left.\xi \!\!\sim \!(120\,{\rm to}\,140)\right], \\ \!\!{\rm when}\;\; & \!\!m_{\nu_1} \!\sim \!(0\,{\rm to}\,1)\!\times \!10^{-3}\,{\rm eV}, & \!\!m_{\nu_2}\! \sim \!(8.9\,{\rm to}\,9.0)\!\!\times\!\!10^{-3}{\rm eV},\!\! & \!\!m_{\nu_3} \!\sim\! 5.0\!\times \!10^{-2}\,{\rm eV}\,. \\ \\ 
\end{array} 
$$

\ni We can see that for charged leptons $\xi \simeq 0$ and for neutrinos $1/\xi \simeq 0$ (but here they are not exactly 0 in both cases) while for up and for down quarks $\xi \simeq \varepsilon$. In a way, the value of $\xi $ characterizes four kinds of fundamental fermions. 

For charged leptons, in the strict limit of $\xi \rightarrow 0$, we have {\it predicted} one linear relation between $m_e, m_\mu $ and $m_\tau $ (Eq. (20)), giving $m_\tau = 1776.80$ MeV {\it versus} the experimental value $m_{\tau} = 1776.99^{+0.29}_{-0.26}\;{\rm MeV}$, when the experimental values of $m_e$ and $m_\mu $ are used. For neutrinos, in the strict limit of $1/\xi \rightarrow 0$, we have {\it predicted} one linear relation between $m_{\nu_1}, m_{\nu_2} $ and $m_{\nu_3}$ (Eq. (49)), leading to their values (61), when the experimental estimates  $|m^2_{\nu_2} - m^2_{\nu_1}| \sim 8.0\times 10^{-5}\;{\rm eV}^2$ and $|m^2_{\nu_3} - m^2_{\nu_2}|                                                 \sim 2.4\times 10^{-3}\;{\rm eV}^2$ are applied. Then $m^2_{\nu_2} < m^2_{\nu_1} < m^2_{\nu_3}$      {\it i.e.}, the mass ordering of neutrino states 1 and 2 is {\it inverted}, while the position of neutrino state 3 is {\it normal}.

If the simplest form of {\it complementarity} relation works between the neutrino and 
charged-lepton $\xi $'s: $1/\xi^{(\nu)} \sim \xi^{(l)} = 1.8\times 10^{-3}$, then the predictions for neutrinos at the level of two decimals are practically identical to those in the strict limit of $1/\xi^{(\nu)} \rightarrow 0$. 

\vspace{0.3cm}

\ni {\bf 6. Theoretical background}

\vspace{0.3cm}

Finally, the following comment is due. Although the formula (1) has essentially an empirical character, its attractive universality and simplicity can be connected with a speculative background based on a K\"{a}hler-like extension of the Dirac equation ({\it i.e.} on an extension of Dirac's square-root procedure) which the interested reader may find in Ref. [1]. In particular, the  numbers $N_i$ and $\rho_i \;(i=1,2,3)$ being fixed elements of formula (1) can be constructed and  interpreted in this formalism.. In fact, $N_i - 1 = 0,2,4$ is there the number of {\it additional} bispinor indices appearing in the extended Dirac equation and obeying Fermi statistics (along with Pauli principle) that enforces their antisymmetrization and so, restricts to zero the related additional spin.  This antisymmetrization is also the reason, why there are precisely {\it three} \SM fermion generations {\it i.e.}, $N_i - 1 = 0,2,4$ (since any of $N_i -1$ additional bispinor indices can assume four values, what then implies that $N_i -1 \leq 4$). Thus, an analogue of Pauli principle works here {\it intrinsically} for $N_i -1$ additional bispinor indices treated as physical objects. Besides, there is one bispinor index, distinguishable from those additional, that must be associated with the set of $SU(3)\times SU(2)\times U(1)$ labels numerating 15+1 states in each \SM generation. So, there are altogether $N_i = 1,3,5 $ bispinor indices treated as physical objects that can be called {\it algebraic spin-1/2 partons} of leptons and quarks. The other fixed elements of formula (1), the generation-weighting factors $\rho_i = 1/29, 4/29, 24/29 $, when multiplied by 29, tell us {\it how many times} the lepton or quark wave-functions of three generations {\it are realized} (up to the factor $\pm1$) in the extended Dirac equation.

The above picture of leptons and quarks is {\it intrinsic} in the sense that the introduced spin-1/2 partons are entirely algebraic objects, not related to any individual positions within leptons and quarks. Such a picture of algebraic compositness abstracts from the spatial notion of familiar compositness and remains in an ideological analogy to the Dirac intrinsic picture of spin 1/2 that abstracts from the spatial notion of familiar angular orbital momentum. In our opinion, this picture of leptons and quarks is especially attractive in its {\it pure} form characterized above. It may happen, however, that in reality the bispinor indices appearing in this picture show up only the summit of a hidden iceberg consisting of more familiar {\it spatial spin-1/2 partons} bound (usually) in orbital S states.

Of course, in spite of some hints, the proposed fundamental-fermion mass formula (1) cannot be justified on the ground of the present particle theory that, strictly speaking, does not include any predictive mass theory. The role of this empirical mass formula is just to help in finding the adequate mass theory, in particular in the neutrino sector. Similarly, the empirical Balmer formula for hydrogen spectrum (1885) could not be justified on the ground of prequantal theory but, nevertheless, it inspired the discovery of Bohr semiclassical quantum conditions and, in consequence, of quantum mechanics by Schr\"{o}dinger and Heisenberg.

As far as the problem of uniqueness in the deduction of spectral formulae from experimental data is concerned, the situation of the empirical mass formula (1), especially in the neutrino sector, is evidently much worse than that of the Balmer formula which could be derived from very complete and precise spectroscopic data already available then for hydrogen. In fact, in spite of the enormous progress in neutrino physics, data on neutrino masses are still rather imprecise. In particular, for the mass $m_{\nu_1}$ only upper limits are estimated from tritium $\beta$ decay experiments or astrophysical observations, say, $m_{\nu_1} < 2.3$ eV or $m_{\nu_1} < 0.6$ eV, respectively, although $|\Delta m^2_{21}|$ and $|\Delta m^2_{32}|$ are reasonably well known. At any rate, we can expect a growing improvement in experimental estimation of $|\Delta m^2_{21}|$ and $|\Delta m^2_{32}|$. Generally, the neutrino physics is in a state of rapid development, and various new results and surprises are possible.

I am indebted to Stefan Pokorski for a stimulating discussion.

\vfill\eject

~~~~
\vspace{0.5cm}

{\centerline{\bf References}}

\vspace{0.5cm}

{\everypar={\hangindent=0.6truecm}
\parindent=0pt\frenchspacing

{\everypar={\hangindent=0.6truecm}
\parindent=0pt\frenchspacing

[1]~ W. Kr\'{o}likowski, {\it Acta Phys. Pol.} {\bf B 33}, 2559 (2002) [{\tt hep--ph/0203107}]; {\tt hep--ph/0504256}; and references therein.

\vspace{0.2cm}

[2]~W. Kr\'{o}likowski, {\it Acta Phys. Pol.} {\bf B 35}, 673 (2004).

\vspace{0.2cm}

[3]~W. Kr\'{o}likowski, {\it Acta Phys. Pol.} {\bf B 36}, 2051 (2005) [{\tt hep--ph/0503074}]; {\tt hep--ph/0504256}.

\vspace{0.2cm}

[4]~Particle Data Group, {\it Review of Particle Physics, Phys.~Lett.} {\bf B 592} (2004).

\vspace{0.2cm}

[5]~{\it Cf. e.g.} G.L. Fogli, E. Lisi, A. Marrone and A. Palazzo, {\tt hep--ph/0506083}.

\vspace{0.2cm}

\vfill\eject

\end{document}